# Properties of Dark Matter Revealed by Astrometric Measurements of the Milky Way and Local Galaxies

Edward Shaya[1], Robert Olling,
Massimo Ricotti, Stuart Vogel
University of Maryland
and
Steven R. Majewski,
Richard J. Patterson
University of Virginia
and
Ron Allen,
Roeland P. van der Marel
STScI
and
Warren Brown
CfA/Harvard U.
and
James Bullock
UC, Irvine
and
Andreas Burkert
University of Munich
and
Francoise Combes
LERMA, Observatoire de Paris

and
Oleg Gnedin
University of Michigan
and
Carl Grillmair, Shri Kulkarni
Caltech
and
Puragra Guhathakurta
UC, Santa Cruz
and
Amina Helmi
Kapteyn Astronomical Institute
and
Kathryn Johnston
Columbia University
and
Pavel Kroupa
University of Bonn
and
George Lake, Ben Moore
University of Zurich
And
R. Brent Tully
U. of Hawaii

Figure 1 - Simulation of Local Group in Formation from simulations by B. Moore et al. (2001), Physical Review D64, 063508

[1] – Astronomy Dept., College Park, MD 20742, eshaya@umd.edu, phone – 301-405-2040

# Abstract

The fact that dark matter (DM), thus far, has revealed itself only on scales of galaxies and larger, again thrusts onto astrophysics the opportunity and the responsibility to confront the age old mystery "*What is the nature of matter*?" By deriving basic data on the nature of DM – e.g., mass of its particle(s), present mean temperature, distribution in galaxies and other structures in the universe, and capacity for dissipational collapse – we will be uncovering the properties of the dominant species of matter in the universe and significantly extending the standard models of particle physics. Determining the mass of the DM particle to an order of magnitude would help to sort out the particle family to which it (or they) belongs. Beyond mass, there are issues of stability. The DM particle may be unstable with a measurable half-life, or it may become unstable after absorbing a certain amount of energy from collisions. In both cases it would contribute to the present hot dark matter component.

Some key parameters of DM can most accurately be measured in the very nearby universe because DM dominates the mass in the outer Milky Way (MW), in other galaxies in the Local Group, and in the Local Group in its entirety. The presence and amount of DM can be quantified by study of dynamical processes observable in fine detail within these entities. Precise measurements of 3-D velocities for stars, coherent star streams, and stars in satellite stellar systems out to the edge of the Galaxy can reveal "*what is the shape, orientation, density law, and lumpiness of the dark matter halo*" as well as "*what is the total mass of the Galaxy*?" Similarly, 3-D velocities of galaxies in the Local Group can reveal "*what are the masses of individual dominant galaxies, the Local Group in total, and the density of the more smoothly distributed warm and hot DM*?"

The advent of large aperture, ground-based telescopes now makes the measurement of radial velocities for faint stars routine, but radial velocities enable only less sensitive, statistical analyses of global dynamics. Deriving the much more challenging transverse positions of faint, distant stars requires both accurate proper motions and distances, which can only be derived with advanced technologies for astrometric measurements. NASA's SIM Lite is a targeted astrometric interferometer with an unparalleled end-of-mission accuracy of 4 µas in position or 2.5 µas yr$^{-1}$ (~1 µas yr$^{-1}$, if extended to 10 years) in proper motion when used in wide-angle mode. The advanced technologies that make SIM Lite possible have now been successfully developed and the mission is ready for implementation. SIM Lite represents a unique and ideal next step in the investigation of DM – just one of several themes in which it could make major breakthroughs.

## I. Introduction

In the outer regions of galaxies, rotation curves from 21-cm emission are generally flat rather than falling as had been expected. Thus, while the light from stars falls away nearly exponentially with radius from the center, the enclosed mass is apparently growing roughly linearly. The dramatic rise in mass-to-light ratio is strong evidence that an invisible species of matter lurks in the halos of galaxies. In addition, there is long standing evidence that the velocity dispersion in galaxy clusters implies a much greater mass than would be inferred from their individual galaxies. This is dramatically confirmed by both the mass required for hydrostatic equilibrium of X-ray emitting gas and by weak gravitational lensing by galaxy clusters. From COBE and WMAP, we know that only 4% of the mass-energy density in the universe is baryonic. Of the remainder, about 22% is composed of dark matter, non-baryonic material that does not interact with visible matter, but whose presence can be inferred from its gravitational effect on the dynamics of visible matter.

The existence of ubiquitous DM has been accepted for years, but with few hard facts about its properties.   The total mass of every galaxy, galaxy group and galaxy cluster depends critically on where mass growth ends, but this is generally not observable.  Just beyond the outermost measured 21-cm or X-ray isophotes of galaxies the DM distribution simply becomes unknown. While weak lensing provides some rough constraints through statistical ensemble averaging, there has not yet been a reliable total mass or mass distribution for any galaxy or group of galaxies to understand DM on such scales.

The concordance $\Lambda$ Cold Dark Matter ($\Lambda$CDM) model for the formation of structure in the Universe, while remarkably successful at describing observations of structure on large scales, continues to be challenged by observations on galactic scales.  Fortunately, CDM models and their various proposed alternatives make a variety of testable predictions that turn the Local Group and the MW into key laboratories for exploring DM.  In this context, aspects of various well-known discrepancies between CDM theory and observations, such as the "missing satellites problem", the "central cusps problem" and problems with the angular momentum distributions in galaxies are ripe for exploration in our local universe.  In particular, high accuracy, µ-arcsecond astrometric observations would allow definitive tests of dynamical effects specifically predicted by CDM models and allow a means to determine the present spatial distribution and primordial phase space distribution of DM in the Local Group.

Some tests of local DM fall within the region of the MW accessible to the ESA's Gaia mission, while several key experiments can only be carried out definitively within a part of the (apparent magnitude)-(astrometric precision) parameter space that is beyond the reach of the Gaia sky survey.  To study dynamics in the outer parts of the MW and in the Local Group requires building an observatory that can carry out pointed observations to reach faint targets with greater precision per visit.  An instrument with the expected capabilities of SIM could strongly constrain "*what is the lowest mass object of the substructure mass function*" and ultimately answer the fundamental question, "*What is the mass of the DM particle?*"

## II.    Science Questions to be Addressed

### A.  The Origin of Extreme Dwarf Galaxies

Our understanding of the satellite galaxy population has been revolutionized in recent years thanks to an increasingly deep census of satellite galaxies around the MW and M31 (*e.g*., Willman et al. 2005; Martin et al. 2006; Zucker et al. 2006; Belokurov et al. 2007; Majewski et al. 2007).  Indeed, the number of known Local Group satellite galaxies has more than doubled since 2005.  It remains to be seen how these new galaxies are to be interpreted in the context of the missing satellites problem.  Understanding the origin of galaxies as extreme as the ultra-faint dwarfs (e.g., Boo II with L ~ 1000 $L_\odot$; Walsh et al. 2008) represents a fundamental challenge because we do not know whether to extend existing galaxy formation models to these extremes or to appeal to tidal interactions to explain their structure and dynamics. Only SIM Lite can measure the proper motions of even the brightest stars in these diminutive stellar systems, but such data are needed to derive the orbits of these systems and determine the degree to which these extreme objects were shaped by past encounters with the MW.

## B. The Shape of the Milky Way DM Halo from Tidal Streams

ΛCDM predicts that DM halos of MW sized galaxies are typically triaxial but become rounder at large radii (Allgood et al. 2006; Macciò, Dutton, & van den Bosch 2008). Current constraints on the shape of the MW DM halo from modeling the debris of the Sagittarius dwarf spheroidal depend on how the extant data are analyzed, yielding shapes from prolate (Helmi 2004), to nearly spherical (Fellhauer et al. 2006), to oblate (Law, Johnston, & Majewski 2005; Martinez-Delgado et al. 2004). Micro-arcsecond astrometry of stars in the orbits of tidal streams would provide more definitive measurements of the halo shape, orientation, and mass profile with radius. These measurements will determine how the shape of the inner Galactic halo has been influenced by dissipation and the formation of the Galactic disk (Kazantzidis et al. 2004; Zentner et al. 2005).

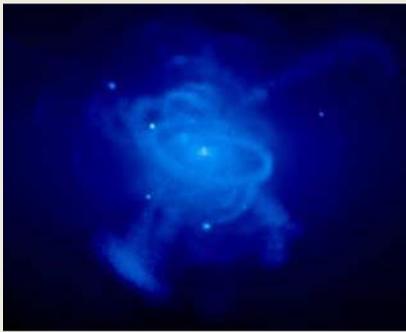

Figure 2 - Surface brightness map of a simulated stellar halo formed within a ΛCDM context (Bullock & Johnston 2005). The image is 300 kpc on each side, and shows just the stellar halo component

Models of structure formation predict and recent observations confirm that the luminous Galactic halo is inhomogeneous and coursed by many streams of debris tidally pulled out of accreted satellites. Once precision proper motion measurements are combined with observed angular positions, radial velocities, and estimated distances, stream stars will provide a uniquely sensitive probe of the global distribution of mass in the MW. Starting from its full phase-space position, each stellar orbit can be integrated backwards within some assumed Galactic potential. Only in the correct potential will the stream stars ever coincide in position and velocity with the parent satellite and each other. Tests of this idea with simulated data observed with precisions of 10 μas yr$^{-1}$ proper-motions and 1 km s$^{-1}$ radial velocities suggests that 1% accuracies on Galactic parameters (such as the flattening of the gravitational potential and the circular speed at the Solar Circle) can be achieved with tidal tail samples as small as 100 stars (Johnston et al. 1999).

Applying this method to several streams at a variety of Galactocentric distances and orientations with respect to the Galactic disk would allow us to build a comprehensive map of the distribution of DM in the halo. In the last few years evidence for dSph galaxy tidal tails has been discovered around MW satellites at very large Galactic radii (Muñoz et al. 2006; Muñoz, Majewski, & Johnston 2008; Sohn et al. 2007), and other distant streams are known (e.g., Newberg et al. 2003; Pakzad et al. 2004; Clewley et al. 2005). If $V$ = 18-19$^{th}$ mag stars are observed with μ-arcsecond precision, such distant streams can be used to trace the Galactic mass distribution as far out as ~200 kpc with an unprecedented level of detail. This would provide the first, accurate 3-D observational assessment of the shape of a galactic-scale DM halo.

Stars in tidal tails also promise to place limits on the possible existence of a large population of long captured small halos or satellites. Large-mass lumps in the MW's potential should scatter stream stars and reduce their coherence (Ibata et al. 2002; Johnston, Spergel, & Haydn 2002; Mayer et al. 2002). For this experiment, the coldest streams (*e.g.*, from globular clusters like Palomar 5 — Grillmair & Dionatos 2006) would be most sensitive. Early tests of such scatterings using only radial velocities of the Sagittarius stream suggest a MW halo smoother than predicted (Majewski et al. 2004), but this result is based on debris from a satellite with an already sizable intrinsic velocity dispersion, therefore the constraint is not strong.

### C. Probing Galaxy Potentials with Hypervelocity Stars

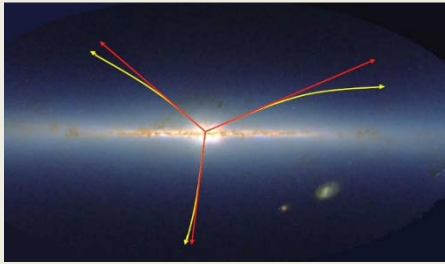

Figure 3 - Schematic representation of the trajectories of hypervelocity stars. Their 3-dimensional positions and velocities trace the non-spherical shape of the Galactic potential.

A complementary method for sensing the shape of the Galactic potential could come from proper motions accurate to a few $\mu$as yr$^{-1}$ of hypervelocity stars (HVSs; Gnedin et al. 2005). HVS stars could be ejected at speeds exceeding 1000 km s$^{-1}$ after the disruption of a close binary star system deep in the potential well of a massive black hole (Hills, 1988) or by the interaction of a single star with a binary black hole (Yu & Tremaine 2003). HVSs are valuable tools because they are found in all directions on the sky and probe the DM halo to large depths (>100 kpc). Recently Brown et al. (2006) reported on five stars with Galactocentric velocities of 550 to 720 km s$^{-1}$, and argued persuasively that these extreme velocity stars can only be explained by dynamical ejections associated with a massive black hole. The total count of HVSs is now 14 (Brown, Geller & Kenyon 2009). After the success of these initial surveys to find HVSs, it is likely that many more will be discovered in the next few years. If these stars indeed come from the Galactic center, the orbits are tightly constrained by knowing their point of origin. In this case the non-spherical shape of the Galactic potential – due in part to the flattened disk and in part to the triaxial dark halo – will induce non-radial inflections (which will be primarily transverse to the line-of-sight at large radii) in the velocities of the HVSs. Distant HVSs provide strong constraints on the DM halo, but will be faint and have small proper motions (<0.1 mas/yr), therefore SIM-like capabilities are required.

### D. Dark Matter within Dwarf Galaxies

Based on internal dynamics and velocity dispersions of groups of dwarf galaxies, dwarf spheroidal (dSph) galaxies occupy the least massive known DM halos in the Universe. Dwarf spheroidals are also unique among all classes of galaxies in their ability to probe the particle nature of DM, because phase-space cores resulting from the properties of the DM particle are expected to be most prominent in these small halos (Tremaine & Gunn 1979; Hogan & Dalcanton 2000). In recent years, the measurement of line-of-sight velocities for upwards of a thousand stars in several dSphs has allowed for determinations of their masses (*e.g.*, Strigari et al 2008; Walker et al. 2007). However, despite the great progress made in estimating masses of these systems, determining the logarithmic slopes of their central density profiles, and thus the nature of the DM contained within, remain elusive. Modeling of the mass distribution of the dSphs requires a solution of the equilibrium Jeans equation, which gives a relation between the components of the intrinsic velocity ellipsoid and the mass density profile of the DM halo. The observed velocity dispersion is 1-dimensional, and thus uncertainty arises in the slope of the DM density profile because the ratio of the radial and tangential components of the velocity variance is not known. Even with 1000 line-of-sight velocities, the relative error on the log-slope at the King core radius is ≈ 1. Proper motion measurements of 10 µas yr$^{-1}$ precision per 16$^{th}$-18$^{th}$ mag stars in nearby dSph (~100 kpc) provide ~ 5 km s$^{-1}$ velocity measurements in the 2 transverse directions, which would be sufficient to determine the velocity ellipsoid as a function of radius and discriminate between cored and cusped density profiles (Strigari et al 2007).

### E. Dark Matter and Dynamics on the Galaxy Group Scale

We are finally sufficiently technologically advanced to begin observing proper motions of the many nearby galaxies to solve for the orbits and masses of individual galaxies. If one measures the proper motions of nearby galaxies with global accuracies of a few µas yr$^{-1}$, one would have another pair of high quality phase-space components with which to construct flow models and determine histories and masses for galaxies and galaxy groups. For a galaxy 1 Mpc away, 4 µas yr$^{-1}$ corresponds to 19 km s$^{-1}$ transverse motion, compared to the expected transverse motions in the field, ~100 km s$^{-1}$. The low velocity dispersion of stars in a typical dwarf galaxy ensures that after averaging just a few random stars, the contribution to the error from the internal motions would be negligible. For larger galaxies, rotation models, adjusted to the observed velocity profiles or direct measurements of the rotation from astrometry, can be removed from the motions to deliver better than 20 km s$^{-1}$ accuracy.

Within 5 Mpc, there are 27 galaxies known to have stars brighter than 20$^{th}$ mag. Measurements of the 3-D motion of these galaxies would be of lasting importance in modeling of the formation of the Local Group, several nearby groups, and the local region of the Local Supercluster. Although the brightest stars in galaxies beyond the MW satellite system are > 16$^{th}$ mag, Gaia will probably be able to obtain proper motion vectors for M31 and M33 after averaging because, for these 2 rich systems, it can access sufficient numbers of star, but these measurements would be limited by correlated errors in the Gaia reference grid. It will take SIM to obtain the proper motions of the other galaxies in the vicinity to sufficient accuracy.

There are two competitive techniques to analyze the expected proper motion data of nearby galaxies, constrained N-body and Least Action methods. Both techniques allow one to solve for the trajectories that resulted in the present distribution of galaxies and the individual total masses. In the constrained N-body technique, an iterative cycle is set up in which initial perturbations are altered to improve a comparison between present galaxy positions and velocities and those derived from evolving the perturbations forward in time with an N-body code (Klypin et al 2003). The Least Action method (LA, Peebles 1989, 2001, Shaya et al 1995) technique makes use of the constraint that early-time peculiar velocities were small. The numerical problem becomes a mixed boundary value problem with peculiar velocity constraints at early times and positional constraints at late times. The masses are altered to improve the fit between observed and predicted proper motions. The Least Action Method is easier to apply, while the constrained N-body provides a better representation of the effects of late-time infall.

Complex orbits are rare when solving for orbits of galaxies separated by >> 100 kpc from any major galaxy, therefore the uncertainty associated with the number of revolutions since early times is not a concern. With knowledge of the proper motions of a few dozen galaxies, the problem becomes fully constrained, and one can solve for the total masses of the dominant galaxies out to 6 Mpc and for the uniformly distributed (hot) component.

### III. Needed Science Data and Mission Requirements

The detailed study of DM at the outer parts of the Galaxy, within the satellites of the Galaxy, and throughout the Local Group can be accomplished with high precision astrometry consisting roughly of 10 µas global positioning per epoch over a 5 year baseline resulting in about 3 µas yr$^{-1}$ proper motions with respect to the extragalactic reference frame. Supergiants can be seen throughout the Local Group with $V < 20^{th}$ magnitude and giants can be used throughout the MW down to about the same brightness. To do this requires a reference grid, tied to the extragalactic sources, at about this precision for several thousand stars, along with the ability to refer targets accurately to grid stars over about a 15 degree field of view.

### IV. Recommendation

Using conventional instruments (imaging/spectroscopy) has led to progress in understanding DM, but it has been particularly slow and limited. Much greater progress in the future is likely using specialized astrometric instruments/telescopes to provide the highest precision measurements of Galactic and even extragalactic dynamics. Astrometry on future huge ground-based telescopes, such as GMT and TMT, using adoptive optics and background galaxies in the image as a reference frame is an intriguing possibility, but their performance remains wholly unknown and, realistically, these telescopes will not be in operation until the next decade

NASA's SIM Lite is a targeted precision astrometric telescope with single-measurement accuracy of 1.0 µas (narrow-angle mode) and accuracy for minimum detectable astrometric signature of 0.21 µas, from magnitude - 1.5 to 20 (Unwin et al 2008). In the wide-angle mode, SIM Lite has an end-of-mission accuracy of 4 µas or 2.5 µas yr$^{-1}$ (~1 µas yr$^{-1}$, if extended to 10 years). SIM began as a development program in the early 1990s and has been developed as a launchable instrument. Starting in 1996, the instrument's performance has been verified by sophisticated test-beds that simulate a flight-like environment. In 2005 the technology program was completed and officially signed-off through a series of independent peer reviews. In short, SIM is largely finished with formulation and ready for implementation. The detailed exploration of DM is just one of several themes in which this mission could make major breakthroughs.

### V. References and Further Reading

- Allgood, B., et al., *MNRAS* 367 (2006): 1781-1796.
- Belokurov, V., et al., *ApJ* 658, (2007):337-344.